Цифровая экономика

# ИНСТИТУЦИОНАЛИЗАЦИЯ ЦИФРОВОЙ ТОРГОВЛИ В РОССИЙСКОЙ ФЕДЕРАЦИИ: ОБРАТНЫЙ ОТСЧЕТ

**Калужский Михаил Леонидович**

*Кандидат философских наук, доцент*
*МОФ «Фонд региональной стратегии развития», исполнительный директор*
*Омский государственный технический университет, каф. «Организация и управление наукоемкими производствами», доцент*
*Омск, Российская Федерация*
*frsr@inbox.ru*

**Аннотация**

*Институционализация цифровой торговли является одним из важнейших направлений формирования информационного общества в Российской Федерации. Исследования отражают наметившееся отставание российского индекса готовности экономики поддерживать онлайн-покупки. Автор анализирует причины отставания в контексте институциональных особенностей развития цифровой торговли. В качестве основного препятствия, снижающего экономическую эффективность и конкурентоспособность цифровой торговли, выделяется недостаточное внимание государства формированию инновационных институций цифрового рынка.*

**Ключевые слова**

*сетевая экономика; цифровая торговля; цифровой рынок; электронная коммерция; институциональная политика; контрактное производство; сетевое предпринимательство; маркетплейсы; логистический провайдинг*

## Введение

Цифровизация не просто определяет ключевое направление развития российской экономики, но служит источником институционального роста цифровой торговли. Указом Президента РФ № 203 от 09.05.2017 г. определены национальные интересы, затрагивающие сферу цифровой торговли, среди которых следует выделить:

1) формирование виртуальных рынков и обеспечение лидерства на них за счет развития российской экосистемы цифровой экономики;

2) обеспечение недискриминационного доступа к товарам и услугам российских поставщиков;

3) поддержка отраслей, использующих преимущества информационных технологий;

4) увеличение экспорта за рубеж несырьевых товаров и услуг;

5) создание платежной и логистической инфраструктуры интернет-торговли.

Перед Правительством РФ поставлена задача формирования технологической основы цифровой экономики, в том числе через повышение доступности электронных форм коммерческих отношений предприятиям малого и среднего бизнеса [1]. Для ее решения приняты и реализуются «Стратегия развития информационного общества в Российской Федерации на 2017-2030 годы», национальная программа «Цифровая экономика Российской Федерации» и федеральные проекты «Информационная инфраструктура», «Цифровые технологии», «Цифровые услуги и сервисы онлайн» и др.

## 1 Российская цифровая торговля в мировых рейтингах

Рейтинговые оценки цифровой торговли в России свидетельствуют о том, что пока рано говорить о значительных успехах. Согласно аналитическому отчету «Интернет-торговля в России: 2021» компании *Data Insight* институциональное и инфраструктурное развитие цифровой торговли в Россия пока далеко от совершенства (см. табл. 1) [2, с. 26].

*Таблица 1. Рейтинг Российской Федерации в мировой системе цифровой торговли*

| Рейтинг | Место |
|---|---|
| *Best Countries For Investment In E-commerce And Digital Sector (Ceoworld)* – индекс привлекательности страны для инвестирования в электронную коммерцию и цифровой сектор | 15 |
| *The Inclusive Internet Index* – индекс доступности цифровой инфраструктуры, цен, локального контента, вовлеченности пользователей и культурных факторов | 25 |
| *The Ease of Doing Business Index* – индекс благоприятности условий предпринимательской деятельности | 28 |
| *UNCTAD B2C E-commerce Index Ranking (UNCTAD)* – индекс готовности экономики поддерживать онлайн-покупки | 41 |

В целом приведенный рейтинг довольно наглядно отражает сложившееся положение в цифровой торговле. Индекс *Ceoworld* показывает лучший результат за счет доминирования на рынке крупных торговых сетей, интернет-магазинов и маркетплейсов. *Inclusive Internet Index* демонстрирует вовлеченность пользователей в среду цифровой торговли. *Ease of Doing Business Index* отражает отставание предложения отечественных продавцов от покупательского спроса.

Хуже всех выглядит индекс *UNCTAD* (ООН), согласно которому в 2020 г. готовность экономических институтов поддерживать онлайн-покупки в России находилась на 41 месте из 152 стран мира [3, с. 14]. Именно этот индекс отражает недостаточную эффективность институциональной политики государства в сфере цифровой торговли.

## 2 Институциональные процессы в цифровой торговле

Цифровая торговля являет собой типичный пример технологической инновации, выступающей следствием очередного институционального цикла [4, с. 25]. Институциональный цикл проходит в своем развитии те же этапы, что и любой жизненный цикл в экономике, менеджменте или маркетинге: выход на рынок, рост, зрелость и упадок. На первых этапах институционального цикла доминируют институции (поведенческие шаблоны и традиции), спонтанно возникающие в рыночной среде за пределами влияния государства. На последних этапах государство регламентирует и ставит под свой контроль экономическую активность. Этот процесс и называется институционализацией.

Развитие институций всегда опережает развитие институтов, поскольку они возникают вследствие экономической активности рыночных субъектов, пытающихся выжить под гнетом рыночных доминантов и государства. Тогда как институты представляют собой результат реактивной деятельности государства на сокращение налогооблагаемой базы вследствие вытеснения традиционных субъектов рынка его неинституционализированными игроками. Проще говоря, государство озаботилось институционализацией цифровой торговли после того, как покупатели стали отворачиваться от традиционных продавцов, а объем отправлений с *AliExpress* превысил объем внутренних отправлений через ФГУП «Почта России».

Поэтому не следует ожидать синхронного развития институтов и институций цифровой торговли. Это противоречило бы самой природе институционального развития. Речь идет о естественном несовершенстве институциональной политики государства и ошибках при определении ее приоритетов. Внедрение экономических новаций неизбежно связано с высокой вероятностью незапланированного поведения рыночных субъектов. Оно нуждается в мониторинге ситуации и корректировке.

## 3 Институциональная эволюция цифровой торговли

В основе институций рынка цифровой торговли лежат конкурентные преимущества ее субъектов, связанные с экономией транзакционных издержек при совершении сделок [5, с. 8]. Цифровая экономика предоставила им неограниченный доступ к аудитории, автоматизацию продаж и сетевую инфраструктуру логистики. Причем, на различных стадиях институционального цикла цифровой торговли указанные преимущества доминируют в определенной последовательности (см. рис. 1).

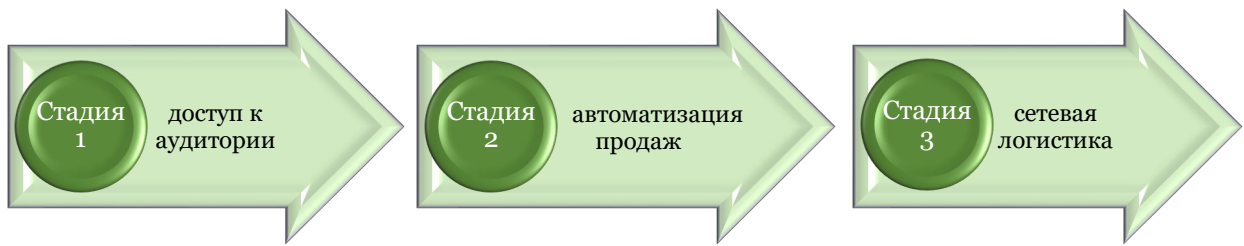

*Рис. 1. Конкурентные преимущества на разных стадиях институционального цикла цифровой торговли*

На первой стадии появилась возможность совершать сделки через электронные доски объявлений, в социальных сетях и на интернет-форумах. Начался бум интернет-магазинов, возникли первые интернет-аукционы, сервисы совместных покупок и дропшиппинг. Структурные изменения происходили вне внимания государства, поскольку на потребительском рынке сделки совершались втемную, налоги с них – не выплачивались. Государственная статистика фиксировала лишь кратное увеличение почтовых отправлений.

На второй стадии начался взрывной рост электронной коммерции. Увеличение масштабов интернет-продаж привело к появлению на рынке провайдеров логистических услуг, сформировавших альтернативную инфраструктуру цифровой торговли. Сильнее всего пострадали традиционные оптово-розничные посредники, кредитно-финансовые организации, а также обеспечиваемые ими налоговые поступления. Государству пришлось приступить к институциональному регулированию цифровой торговли.

Третья стадия знаменуется завершением структурной перестройки экономического ландшафта, доминированием укрупняющихся ключевых игроков рынка и сменой их стратегических приоритетов. Если прежде основными конкурентами субъектов цифровой торговли выступали традиционные оптово-розничные продавцы, то здесь они терпят сокрушительное поражение и вытесняются на задворки рынка. Конкурентная борьба разворачивается за повышение эффективности и оптимизацию бизнес-процессов при возрастающей роли государственного регулирования.

Что будет происходить на четвертой, завершающей стадии институционального цикла цифровой торговли пока трудно предсказать, как и сроки ее начала. Сформируются новые институции и их носители, действующие за рамками институционального регулирования государства. Их появление станет очередной попыткой участников рынка вырваться за рамки удушающего влияния действующих доминантов цифрового рынка. Сейчас до этого еще далеко и на повестке дня стоят совсем иные проблемы.

## 4 Смена институционального вектора

Российская экономика находится в самом начале третьей стадии институционального цикла цифровой торговли, где ведущим фактором конкурентоспособности становится сравнительная эффективность бизнес-процессов [6, с. 334]. Между участниками рынка обостряется конкуренция, сам рынок структурируется, значение государственного регулирования возрастает [7, с. 8]. Кроме того, доминирующая роль в вопросах ассортимента, ценообразования и сбыта переходит от продавцов к потребителям. Они голосуют рублем, и глобализация предоставляет им для этого все возможности.

Институциональная политика государства осложняется новизной стоящих задач. Основная проблема состоит в определении ориентиров и приоритетов институционального строительства. Любая ошибка неизбежно приводит к институциональному тупику, оттоку покупателей и отставанию в развитии цифрового рынка. В результате он переходит под контроль более успешных зарубежных конкурентов [8, с. 113-114]. С другой стороны, наличие успешных конкурентов позволяет изучить их опыт и применить его в своей практике.

Так, к примеру, на непродовольственном рынке цифровой торговли в России доминируют два противоборствующих течения: розничные сети (*Эльдорадо*, *DNS*, *Leroy Merlin* и пр.) и маркетплейсы (*Wildberries*, *Lamoda*, *Ozon*, *Яндекс-Маркет*, *Сбермаркет* и др.). Их противостояние обусловлено разной моделью ведения бизнеса: розничные сети извлекают прибыль из своих продаж, тогда как маркетплейсы получают ее от оказания торговых услуг. Розничные сети стремятся закрыть и защитить каналы сбыта, а маркетплейсы, наоборот, стремятся максимально открыть их.

Первоначально пальма первенства была у розничных сетей, довольно успешно лоббировавших свои интересы через *АКИТ* (Ассоциация компаний интернет-торговли). Их главный интерес состоял в создании институциональных барьеров для неинституционализированной трансграничной торговли. Образно говоря, сделать так, чтобы покупка телефона *Xiaomi* на *Aliexpress* обходилась покупателям столь же дорого, как в России.

Максимальным успехом лоббирования торговых сетей стало снижение порога беспошлинного ввоза товаров для личного пользования до 200 евро. Институциональный эффект такого решения представляется весьма спорным, поскольку поддержку получили не производители отечественной продукции, а сфера торговли. В убытке оказались частные потребители, чья покупательская способность снизилась. При этом институциональный цикл торговых сетей уже находится в начале четвертой стадии: на рынке цифровой торговли взрывной рост продаж показывают не они, а маркетплейсы [9].

Главный интерес маркетплейсов заключается в росте продаж через привлечение максимального числа потребителей, для которых главным фактором является цена товара. Маркетплейсы не извлекают прибыль от продажи товаров – она формируется в процессе оказания логистических услуг продавцам. В отличие от розничных сетей, низкие цены для них не беда, а источник финансирования и институционального роста.

Со сменой рыночного доминанта на глазах меняется и институциональная политика государства. Так, с 28.03.2022 в *ЕАЭС* порог беспошлинного ввоза товаров физическими лицами (временно) возвращен к 1000 евро. Кроме того, лидирующие позиции членов *АКИТ* перешли к крупнейшим маркетплейсам (*Wildberries*, *Ozon*, *Avito*, *Lamoda*, *Яндекс-Маркет*), что не могло не сказаться на смене приоритетов ее лоббистской деятельности.[1] Вектор институционального развития цифровой торговли в России сменился от попыток ограничить свободу неинституционализованных ее участников к формированию ориентированной на них рыночной инфраструктуры.

## 5 Большой разворот

Роль маркетплейсов на рынке цифровой торговли трудно переоценить. Первоначально источником их институционального роста был переток покупателей из традиционной торговли с более высоким уровнем трансакционных издержек и розничных цен. Однако к концу 2010-х гг. этот ресурс исчерпал себя. Сегодня на потребительском рынке сопротивление им оказывают розничные сети, с разной степенью успешности осваивающие цифровую торговлю. Наиболее эффективным оружием против них является снижение розничных цен и повышение доступности товаров для покупателей.

2021 год ознаменовался тектоническими изменениями ценовой политики ведущих российских маркетплейсов: они кратно снизили комиссию для продавцов. Снижение комиссии маркетплейсов составило у *Wildberries* до 5-15% (было 38%), у *Ozon.ru* до 5-8% (было 5-25%), у *Яндекс-Маркета* до 2% (было 3-20%). Правильность принятого решения подтвердилось феноменальным приростом продаж (в 2-3 раза) (см. табл. 2) [10].

---

[1] Стандарты качества / Бизнесу // АКИТ. URL: https://akit.ru/business/standards (дата обращения 12.06.2022).

*Таблица 2. Результаты лидирующих маркетплейсов России за 2021 г.*

| Маркетплейс | онлайн-продажи | | заказы | | средний чек | |
|---|---|---|---|---|---|---|
| | млрд. руб. | прирост | млн. шт. | прирост | руб. | прирост |
| Wildberries | 805,7 | +95% | 771,9 | +153% | 1040 | -23% |
| Ozon | 446,7 | +126% | 221,2 | +199% | 2020 | -24% |
| Яндекс-Маркет | 132,6 | +180% | 29,7 | +151% | 4110 | +12% |
| AliExpress | 106,1 | +116% | 48 | +152% | 2210 | -14% |
| Lamoda | 71,2 | +34% | 14,1 | +15% | 5050 | +17% |

При этом прирост продаж был обратно пропорционален размеру среднего чека: чем меньше сумма покупки, тем больше желающих ее совершить. Из общей картины несколько выбивается *Яндекс-Маркет*, но только за счет широкого присутствия на нем розничных сетей, наоборот, ориентированных на прирост среднего чека.

Особняком на этом фоне стоит маркетплейс *Lamoda*, демонстративно игнорирующий институциональные тренды цифрового рынка. У него самые высокие тарифы – 35-70% от розничной цены продаваемых товаров. Можно предположить, что его прибыль значительно превосходит прибыль продавцов товара. В этом *Lamoda* похож на розничные сети. Неудивительно, что прирост его показателей стабильно ниже прироста продаж других лидеров цифрового рынка.

Следует особо отметить наличие огромного потенциала продаж, связанного со снижением суммы среднего чека. Все это задает тренд на совершенствование логистических технологий, направленных на уменьшение транзакционных издержек и снижение розничных цен для покупателей. Участники цифровой торговли, действующие в рамках указанного тренда, добиваются наилучших результатов.

## 6 Новые горизонты цифровой торговли

Практика показывает, что институции цифровой торговли оказывают решающее влияние на конкурентоспособность ее субъектов [11, с. 60-61]. Отставание *E-commerce Index Ranking* лишь подтверждает необходимость корректировки институционального регулирования цифровой торговли и смены его приоритетов. Игнорирование рыночных трендов и закономерностей резко снижает эффективность государственной политики и конкурентоспособность российской цифровой экономики в целом.

В качестве институциональных ориентиров следует выделить три приоритетных направления развития сетевой экономики и покупательский спрос как движущую силу рыночного механизма. Выделение этих ориентиров связано с наиболее успешными институциями, определяющими вектор институционального развития цифровой торговли.

1. *Контрактное производство* – институция, основанная на изготовлении продукции независимым производителем по техническому заданию заказчика с отгрузкой «под ключ». Такое производство переходит из категории работ в категорию услуг. Оно не производит собственную продукцию, оказывая предоплаченные услуги заказчикам, что обеспечивает большую экономическую эффективность.

Контрактное производство не нуждается в кредитовании, имеет отрицательную оборачиваемость средств и не несет предпринимательских рисков в торговле. В этой модели инициаторами производства выступают независимые заказчики, отслеживающие конъюнктуру рынка, принимающие на себя предпринимательские риски и финансирующие производство за счет собственных средств. Контрактное производство становится придатком торговли, ориентирующейся на потребительский спрос.

*Пример*: Биржа контрактного производства Московского инновационного кластера.[2]

2. *Логистический провайдинг* (англ. *Third Party Logistics*) – институция, основанная на делегировании нестратегических внутрифирменных функций независимым провайдерам логистических услуг. Такой провайдинг также переходит из категории работ в категорию услуг. Провайдеры делятся с заказчиками экономией на масштабе оказываемых услуг, за счет своей узкой специализации обеспечивая более высокое качество и эффективность.

---

[2] Биржа контрактного производства // Московский инновационный кластер. URL: https://i.moscow/contract_exchange (дата обращения: 18.06.2022).

Практически любая внутрифирменная функция может быть передана независимому провайдеру: бухгалтерский учет, обработка заказов, разработка технической документации, организация продаж, документооборота и т.д. [12, с. 57]. Высший уровень логистического провайдинга (5PL) предполагает делегирование как функции, так и контроля за ее реализацией по принципу «передал и забыл». В идеальной модели заказчик сосредотачивается на стратегическом направлении деятельности, а все сопутствующие функции делегирует внешним провайдерам.

*Примеры*: маркетплейсы, аутсорсинговые и фулфилментовые компании, бухгалтерские сервисы и т.д.

3. *Сетевое предпринимательство* – институция, основанная на использовании преимуществ виртуальной среды, сетевой экономики и цифровой торговли. Они позволяют сократить затратность ведения бизнеса и снижают входной барьер для участников цифрового рынка, сокращая временные затраты на реализацию бизнес-проектов.

Идеальная модель сетевого предпринимательства стремится к тому, что называется «виртуальная организация», не имеющая ни офиса, ни постоянного штата сотрудников [13, с. 279-281]. Предприниматель здесь выступает в роли организатора и координатора «цепочек создания ценностей», потенциал которых он использует для реализации своего бизнес-проекта. В качестве его сетевых партнеров выступают как контрактные производители, так и провайдеры логистических услуг.

*Пример*: Самодеятельные продавцы маркетплейсов (*Ozon*, *Wildberries* и *Яндекс-Маркет*), продающие контрактные товары под своими брендами.

В своей совокупности все институции образуют экосреду цифровой экономики, в которой цифровая торговля инициирует не только процесс товародвижения, но и товарного производства. Покупатель своим спросом инициирует предпринимательскую активность продавца, который на свой страх и риск организует контрактное производство востребованных товаров и привлекает сетевых провайдеров логистических услуг.

В корне меняются институциональные роли участников сетевого рынка:

*Покупатели* – получают возможность неограниченного выбора, ставя продавцов в условия совершенной конкуренции.

*Продавцы* – откликаются на запросы покупателей, первичный спрос которых инициирует их вторичную предпринимательскую активность.

*Сетевые провайдеры* – оказывают логистические услуги продавцам (не покупателям!), принимая на себя отдельные функции организации товародвижения.

*Производители* – оказывают услуги контрактного производства продавцам, соревнуясь между собой в гибкости производства и скорости выполнения заказов.

Пока наибольшую эффективность показывает институция, в рамках которой покупатель взаимодействует с маркетплейсом, принимающим на себя все заботы по организации товаропотока (*Wildberries*, *Ozon*, *AliExpress*). Однако уже сегодня многие продавцы продают свои товары одновременно на нескольких маркетплейсах, а нелояльные покупатели, сравнивая цены, покупают там, где дешевле. Свобода потребительского выбора размоет диктат маркетплейсов, как они сегодня размывают диктат розничных сетей.

Рано или поздно и маркетплейсы достигнут предела институционального развития и перейдут в категорию «при прочих равных» за счет обострения внутривидовой конкуренции. Если это произойдет, то между продавцами и покупателями сформируется логистическая инфраструктура, в равной мере доступная всем участникам цифрового рынка. Аналогично электричество или компьютеры были когда-то источником рыночной конкурентоспособности, а сегодня воспринимаются как естественная часть рыночного ландшафта.

**Заключение**

Приоритетом институциональной политики государства может стать превращение цифровой торговли в один из локомотивов экономического роста. Для этого необходимо сосредоточиться на снижении транзакционных издержек в сетях товародвижения. В традиционной торговле потребительскими товарами транзакционные издержки (маржа оптово-розничных сетей) составляли 80-100% от конечной цены товара. В маркетплейсах типа Lamoda они и сегодня составляют 30-70% от цены продавца.

Вместе с тем, практика институционального развития цифровой торговли задает совсем иной вектор. Более продвинутые маркетплейсы (*Wildberries*, *Ozon*, *Яндекс-Маркет*) еще в начале 2021 года инициативно снизили размер своей комиссии до 3-5% и это привело к впечатляющим результатам. Так, например, продажи самозанятых на *Wildberries* только в первом квартале 2022 года выросли на 410% (до 2 млрд руб.), а их численность увеличилась почти пятикратно (до 150 тыс. чел.).[3]

В условиях экономического кризиса и западных санкций снижение транзакционных издержек в цифровой торговле способно компенсировать снижение покупательной способности населения. Важно сохранить доступность товаров массового спроса и поддержать товаропроизводителей. Вытесняя из торговой цепочки посредническое звено за счет ускоренной цифровизации торговли, можно не только способствовать решению социальных задач, но и стимулировать рост предпринимательской активности в производственной сфере. Представляется, что именно эта цель должна стать одним из приоритетов институциональной политики государства в отношении цифровой торговли на ближайшие годы.

## Литература

1. Указ Президента РФ № 203 от 09.05.2017 г. «О Стратегии развития информационного общества в Российской Федерации на 2017-2030 годы» / Документы // Президент России. URL: http://kremlin.ru/acts/bank/41919.
2. Интернет-торговля в России 2021: Аналитический отчет. М.: Data Insight, 2022. 156 с.
3. The UNCTAD B2C E-commerce Index 2020 / UNCTAD Technical Notes on ICT for Development 2021 // UNCTAD ONU. 2022. № 17. 22 p.
4. Блуммарт Т. Четвертая промышленная революция и бизнес: как конкурировать и развиваться в эпоху сингулярности. М.: Альпина Паблишер, 2019. 204 с.
5. Данные для лучшей жизни: Обзор доклада о мировом развитии. Washington: Международный банк реконструкции и развития / Всемирный банк, 2021. 39 с.
6. Кочетков Е.П. Цифровая трансформация экономики и технологические революции: вызовы для текущей парадигмы менеджмента и антикризисного управления // Стратегические решения и риск-менеджмент. 2019. Т. 10. № 4. С. 330-341. DOI: 10.17747/2618-947X-2019-4-330-341.
7. Антимонопольное регулирование в цифровую эпоху: как защищать конкуренцию в условиях глобализации и четвертой промышленной революции: монография. М.: ВШЭ, 2019. 391 с.
8. Борисова В.В., Юань Х., Тан Л. Стратегии развития электронной платформы Aliexpress в России // Вестник Ростовского государственного экономического университета (РИНХ). 2020. № 4 (72). С. 110-115.
9. Романова Т. Маркетплейсы рвутся вверх: за счет чего выросли обороты крупнейших ретейлеров России / Бизнес // Forbes. [Электронный ресурс]. 7 июня 2022 г. URL: https://www.forbes.ru/biznes/467927-marketplejsy-rvutsa-vverh-za-scet-cego-vyrosli-oboroty-krupnejsih-retejlerov-rossii (дата обращения: 15.06.2022).
10. Рейтинг ТОП-100 крупнейших российских интернет-магазинов. М.: Data Insight, 2022. URL: https://top100.datainsight.ru (дата обращения: 17.06.2022).
11. Слонимская М.А. Сетевые формы организации экономики. Мн.: Беларуская навука, 2018. 279 с.
12. Tan A., Shukkla S. Digital transformation of the supply chain: a practical guide for. Danvers (USA): World Scientific Publishing, 2021. 152 p.
13. Уорнер М., Витцель М. Виртуальные организации. Новые формы ведения бизнеса в XXI веке. М.: Добрая книга, 2005. 296 с.

---

[3] Продажи самозанятых из России на Wildberries выросли на 410% с января 2022 года. 29.06.2022. / Экономика // ТАСС. URL: https://tass.ru/ekonomika/15065193 (дата обращения 13.07.2022).

# INSTITUTIONALIZATION OF DIGITAL TRADE IN THE RUSSIAN FEDERATION: COUNTDOWN

**Kaluzhsky, Mikhail Leonidovich**

*Candidate of philosophical sciences, associate professor*
*Fund of Regional Development Strategy, executive director*
*Omsk State Technical University, department "Organization and management of science-intensive industries", associate professor*
*Omsk, Russian Federation*
*frsr@inbox.ru*

**Abstract**

*The institutionalization of digital trade is one of the most important directions in the formation of the information society in the Russian Federation. The studies reflect the emerging lag in the Russian economy readiness index to support online shopping. The author analyzes the reasons for the lag in the context of the institutional features of the development of digital trade. As the main obstacle that reduces the economic efficiency and competitiveness of digital trade, insufficient attention of the state to the formation of innovative institutions of the digital market is highlighted.*

**Keywords**

*network economy; digital trade; digital market; e-commerce; institutional policy; contract manufacturing; network entrepreneurship; marketplaces; logistics providers*

**References**

1. Ukaz Prezidenta RF № 203 ot 09.05.2017 g. «O Strategii razvitiya informacionnogo obshchestva v Rossijskoj Federacii na 2017-2030 gody» / Dokumenty // Prezident Rossii. URL: http://kremlin.ru/acts/bank/41919.
2. Internet-torgovlya v Rossii 2021: Analiticheskij otchet. M.: Data Insight, 2022. 156 s.
3. The UNCTAD B2C E-commerce Index 2020 / UNCTAD Technical Notes on ICT for Development 2021 // UNCTAD ONU. 2022. № 17. 22 r.
4. Blummart T. Chetvertaya promyshlennaya revolyuciya i biznes: kak konkurirovat' i razvivat'sya v ehpokhu singulyarnosti. M.: Al'pina Publisher, 2019. 204 s.
5. Dannye dlya luchshej zhizni: Obzor doklada o mirovom razvitii. Washington: Mezhdunarodnyj bank rekonstrukcii i razvitiya / Vsemirnyj bank, 2021. 39 s.
6. Kochetkov E.P. Cifrovaya transformaciya ehkonomiki i tekhnologicheskie revolyucii: vyzovy dlya tekushchej paradigmy menedzhmenta i antikrizisnogo upravleniya // Strategicheskie resheniya i risk-menedzhment. 2019. T. 10. № 4. S. 330-341. DOI: 10.17747/2618-947X-2019-4-330-341.
7. Antimonopol'noe regulirovanie v cifrovuyu ehpokhu: kak zashchishchat' konkurenciyu v usloviyakh globalizacii i chetvertoj promyshlennoj revolyucii: monografiya. M.: VSHEH, 2019. 391 s.
8. Borisova V.V., Yuan' KH., Tan L. Strategii razvitiya ehlektronnoj platformy Aliexpress v Rossii // Vestnik Rostovskogo gosudarstvennogo ehkonomicheskogo universiteta (RINKH). 2020. № 4 (72). S. 110-115.
9. Romanova T. Marketplejsy rvutsya vverkh: za schet chego vyrosli oboroty krupnejshikh retejlerov Rossii / Biznes // Forbes. [Ehlektronnyj resurs]. 7 iyunya 2022 g. URL: https://www.forbes.ru/biznes/467927-marketplejsy-rvutsa-vverh-za-scet-cego-vyrosli-oboroty-krupnejsih-retejlerov-rossii (data obrashcheniya: 15.06.2022).
10. Rejting TOP-100 krupnejshikh rossijskikh internet-magazinov. M.: Data Insight, 2022. URL: https://top100.datainsight.ru (data obrashcheniya: 17.06.2022).
11. Slonimskaya M.A. Setevye formy organizacii ehkonomiki. Mn.: Belaruskaya navuka, 2018. 279 s.
12. Tan A., Shukkla S. Digital transformation of the supply chain: a practical guide for. Danvers (USA): World Scientific Publishing, 2021. 152 p.
13. Uorner M., Vitcel' M. Virtual'nye organizacii. Novye formy vedeniya biznesa v XXI veke. M.: Dobraya kniga, 2005. 296 s.